\documentclass[aps,prl,superscriptaddress,floatfix,reprint]{revtex4-1}  

\usepackage{graphicx}
\usepackage{bm}
\usepackage{amssymb}
\usepackage{amsmath}
\usepackage{float}
\usepackage{color}
\usepackage{bbold}

\newcounter{subfigure}[figure]

\newcommand{\bra}[1]{\left\langle #1 \right\vert}
\newcommand{\ket}[1]{\left\vert #1 \right\rangle}

\newcommand{\grad}[1]{\nabla }

\newcommand{\creop}[1]{\hat{\mathrm{ #1 }}^{\dagger}}

\newcommand{\proj}[3]{\ket{#1}_{\!#2}\!\!\bra{#3}}

\begin{document}

\title{Multi-photon quantum interference in a multi-port integrated photonic device}

\author{Benjamin~J.~Metcalf}
\thanks{These authors contributed equally to this work.}
\affiliation{Clarendon Laboratory, University of Oxford, Parks Road, Oxford OX1 3PU, UK}

\author{Nicholas~Thomas-Peter}
\thanks{These authors contributed equally to this work.}
\affiliation{Clarendon Laboratory, University of Oxford, Parks Road, Oxford OX1 3PU, UK}

\author{Justin~B.~Spring}
\affiliation{Clarendon Laboratory, University of Oxford, Parks Road, Oxford OX1 3PU, UK}

\author{Dmytro~Kundys}
\affiliation{Optoelectronics Research Centre, University of Southampton, Southampton, SO17 1BJ, UK}

\author{Matthew~A.~Broome}
\affiliation{Centre for Engineered Quantum Systems and Centre for Quantum Computer and Communication Technology, School of Mathematics and Physics, University of Queensland, 4072 Brisbane, QLD, Australia}

\author{Peter~Humphreys}
\affiliation{Clarendon Laboratory, University of Oxford, Parks Road, Oxford OX1 3PU, UK}

\author{Xian-Min~Jin}
\affiliation{Clarendon Laboratory, University of Oxford, Parks Road, Oxford OX1 3PU, UK}
\affiliation{Centre for Quantum Technologies, National University of Singapore, 117543, Singapore}

\author{Marco~Barbieri}
\affiliation{Clarendon Laboratory, University of Oxford, Parks Road, Oxford OX1 3PU, UK}

\author{W.~Steven~Kolthammer}
\affiliation{Clarendon Laboratory, University of Oxford, Parks Road, Oxford OX1 3PU, UK}

\author{James~C.~Gates}
\affiliation{Optoelectronics Research Centre, University of Southampton, Southampton, SO17 1BJ, UK}

\author{Brian~J.~Smith}
\affiliation{Clarendon Laboratory, University of Oxford, Parks Road, Oxford OX1 3PU, UK}

\author{Nathan~K.~Langford}
\affiliation{Department of Physics, Royal Holloway, University of London, TW20 0EX, UK}

\author{Peter~G.~R.~Smith}
\affiliation{Optoelectronics Research Centre, University of Southampton, Southampton, SO17 1BJ, UK}

\author{Ian~A.~Walmsley}
\affiliation{Clarendon Laboratory, University of Oxford, Parks Road, Oxford OX1 3PU, UK}

\date{\today}

\begin{abstract}
Increasing the complexity of quantum photonic devices is essential for many optical information processing applications to reach a regime beyond what can be classically simulated, and integrated photonics has emerged as a leading platform for achieving this. Here, we demonstrate three-photon quantum operation of an integrated device containing three coupled interferometers, eight spatial modes and many classical and nonclassical interferences. This represents a critical advance over previous complexities and the first on-chip nonclassical interference with more than two photonic inputs. We introduce a new scheme to verify quantum behaviour, using classically characterised device elements and hierarchies of photon correlation functions. We accurately predict the device's quantum behaviour and show operation inconsistent with both classical and bi-separable quantum models. Such methods for verifying multiphoton quantum behaviour are vital for achieving increased circuit complexity. Our experiment paves the way for the next generation of integrated photonic quantum simulation and computing devices.

\end{abstract}

\maketitle
Realizing quantum-enhanced information processors for tasks such as simulation and computation demands experimental systems of sufficient complexity that their dynamics cannot be efficiently determined using classical processors.  Reaching this regime in practice, however, remains a critical open challenge.

Integrated quantum optics provides great promise for enabling photonic experiments, which are otherwise generally limited to relatively small-scale experiments, to reach this new regime of complexity. Chip-based fabrication enables sophisticated networks involving multiple interfering pathways in a compact and stable physical architecture, and pioneering work has shown the viability of this approach for the manipulation of the quantum properties of photons~\cite{PolitiA2008swq, smith2009, Matthews2009, Peruzzo2010,Marshall2009,SansoniL2010pes, Crespi2011,Shadbolt2011}.  To date, experiments have demonstrated up to three-photon higher-order terms from a single nonlinear photon source being coupled into the two input modes of a single interferometer~\cite{Matthews2011}, or alternatively, two-photon correlations in up to 21 waveguide modes~\cite{Peruzzo2010, Owens2011}.  Building a photonic system capable of truly outperforming classical processors, however, can only be achieved by simultaneously increasing the number of modes and interference nodes in the circuit and the number of photons distributed among them.

There are two key outstanding challenges associated with this task of scaling up integrated photonic circuits to these larger systems. First, photon loss exponentially limits the complexity achievable in a quantum circuit, both in terms of the number of circuit elements and the number of photons that can be used effectively.  In integrated photonics, significant losses arise from interfacing the circuit with both sources and detectors and become more pronounced with increased photon numbers~\cite{Thomas-Peter2010a, thomasPeter11NJP}. Ultimately, losses are fundamentally limited by the intrinsic optical properties of the medium and these clearly scale with the circuit size.  Secondly, the monolithic nature of integrated architectures means it is also difficult to verify that the chip meets the design specification.  In particular, it is not possible in general to access individual components \emph{in situ} using the external input and output ports, nor is it always possible to configure ancillary access ports for injecting probe states or performing detection locally.  On the other hand, existing process tomography techniques for verifying the quantum operation of a full chip~\cite{OBrien2004,Shabani2011} become impracticable once it becomes sufficiently complex.  Instead, other simpler ways to measure the chip's transformation are required.

In this work, we demonstrate a critical advance in the complexity of quantum integrated photonic devices by simultaneously increasing the number of photons and the number of spatial modes used.  Using a circuit in which three photons are distributed over eight modes and three coupled interferometers~\cite{ralph2004,lanyon2007,lu2007}, we certify quantum operation beyond both the classical limit and what can be achieved with two photons using a hierarchy of higher-order photon correlation functions.  As part of this, we also provide the first on-chip demonstration of a Hong-Ou-Mandel-type interference effect with three individual input photons. 

We further show that a critical step in verifying the correct quantum operation was characterizing the operating parameters of individual circuit components, and we introduce a simple loss-insensitive method to achieve this using classical light scattered from the device in the transverse direction.  Together with knowledge of the circuit topology, this method allows the full unitary transformation implemented by the device to be reconstructed and used to verify quantum operation. In this paper, we verify three-photon interactions which achieve a complexity sufficient to realise a next generation of on-chip quantum information protocols such as cluster-state generation and teleportation.

\begin{figure}
  \def\svgwidth{1\columnwidth}
  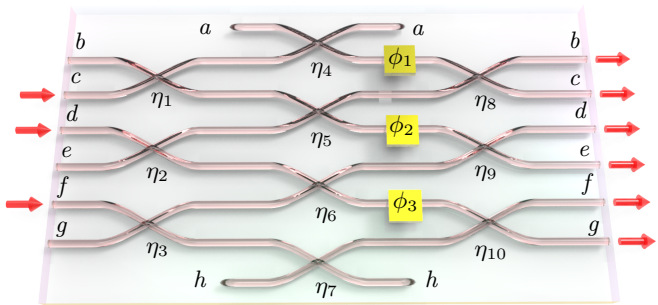
  \caption{\textbf{Schematic of the multi-port waveguide circuit.} The circuit consists of eight modes labelled \emph{a} to
    \emph{h}, ten beam splitters labelled
    $\eta_1$ to $\eta_{10}$, and three variable phase shifters
    $\phi_1$ to $\phi_3$.  Single photons are launched into a subset of \emph{b} to \emph{g} and the output of each mode was monitored by a single-photon detector. The ancillary modes \emph{a} and \emph{h} were not accessible for coupling.}
  \label{fig:chip}
\end{figure}

\section*{Results}
The multiport waveguide circuit used in these experiments consists of three coupled Mach-Zehnder interferometers (MZI) spanning eight spatial modes, with phase control inside the interferometers implemented by thermo-optic phase shifters (see Fig.~\ref{fig:chip} and Methods for a detailed description of the experimental apparatus).  Our main aim is to show genuinely quantum operation of the circuit in a context which demonstrates its full complexity in terms of simultaneously increased number of independent input photons and number of interacting modes and interferometers.  To do this, we inject individual single-photon states into one mode of each of the interferometers (modes \emph{c}, \emph{d} and \emph{f}) and measure the visibility of three-photon nonclassical interference at different combinations of output ports.  If the observed visibility is stronger than that predicted when the single-photon inputs are replaced with classical light, this acts as a witness to the desired quantum behaviour.

To calculate the predicted classical bounds, we must assume that the circuit itself operates completely coherently (unitary operation), so we first characterise the circuit using an element-wise, loss-tolerant approach with classical input states, in the process avoiding any need for resource-intensive quantum process tomography.  We confirm the reliability of our characterisation by first observing two-photon quantum interference and comparing its behaviour with both a classical and a quantum model: the former is clearly inconsistent with the experimental results, while the latter shows good agreement.  Finally, extending this method, we show that the observed three-photon interference measurements exclude with high confidence levels both the classical model, as well as quantum models involving biseparable states.  The nature of the observed three-photon interferences, combined with the known topology of the circuit relative to the input photons, suggest that the results cannot be explained by using a simplified or restricted subsection of the illustrated circuit.  This implies that the experiment utilises the full available complexity of the circuit.

Multiphoton integrated-optics experiments set stringent demands on performance with regard to photon loss~\cite{Thomas-Peter2010a, thomasPeter11NJP}.  Particular care needs to be taken to optimise all experimental efficiencies, especially in experiments utilising down-conversion photon sources, such as this one, to minimise higher-order noise terms~\cite{barbieri2009}.  In order to demonstrate high-brightness multiphoton states ``on-chip'', we have combined a range of technical solutions for optimising the loss properties, including efficient pair-source heralding, optimal coupling of six-channel fiber arrays to the chip and use of a low-loss integrated platform (see Methods). These measures were critical to reaching the level of on-chip complexity achieved in these experiments.

\subsection*{Characterising circuit operation}
The three coupled interferometers in our waveguide circuit, fabricated by UV direct-write technology on a silica-on-silicon  substrate~\cite{smith2009}, involve ten beam splitters and three thermo-optic phase shifters, a circuit which has only previously been realised directly in a simplified polarisation-based encoding with bulk optics~\cite{ralph2004,lanyon2007,lu2007}.  Temperature control by means of a thermo-electric Peltier element maintained stable beam-splitting ratios and phase offsets over many weeks.  Individually characterising these parameters permits us to simulate the required multiphoton interference visibilities.

\begin{figure}
\includegraphics[width=1\columnwidth]{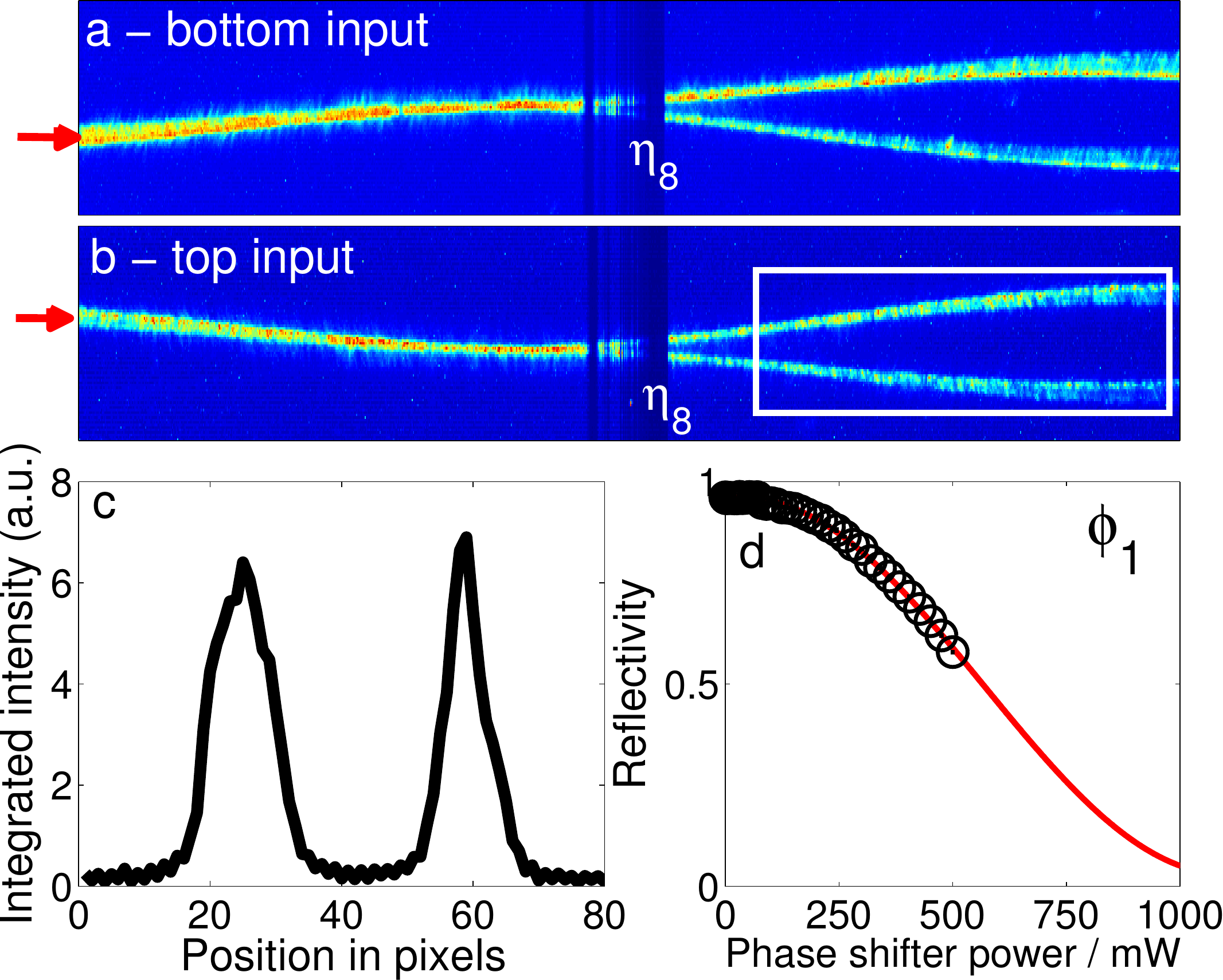}		
	\caption{
	\textbf{Classical characterisation of integrated circuit elements.} \textbf{(a-b)} CCD images of light scattered in the transverse direction from the waveguide. Red arrows denote the direction of propagation of the light. The white box denotes the integration region at the output for the ratiometric analysis. \textbf{(c)} The integrated intensity is used as part of a ratiometeric analysis to determine $\eta_8 = 0.55\pm0.02$. \mbox{\textbf{(d)}} Interference fringe produced by scanning $\phi_1$.  The parameter values obtained from the theoretical fit include the zero-voltage phase offset.}
	\label{fig:charac}
\end{figure}

We measured the beam splitter reflectivities, $\eta_1$ to $\eta_{10}$, by sending continuous-wave laser light through each splitter in turn and calculating the reflectivity using a ratiometric analysis which is independent of coupling and transmission losses~\cite{thomasPeter11NJP}.  This technique uses four measurements for each beam splitter, coupling light in turn into each input port and recording the power at each output port.  Due to the complex circuit topology, it is not generally possible to independently access the input and output modes, so instead, light scattered out of the chip surface was used to measure the output powers.  The ratiometric calculation is insensitive to different scattering efficiencies in the same way as to different interface coupling efficiencies.  Figures~\ref{fig:charac}(a) and (b) show a typical example, with 100\,mW of laser light coupled into the chip (note that splitters 4 and 7 only had one available input, since modes \emph{a} and \emph{h} were not accessible for coupling, see Fig.~\ref{fig:chip}).  The input polarization was adjusted to maximize the amount of the transverse scatter, which was imaged using a CCD with a highly linear response.  Integrating over a specified region then provides the required intensity measurements (Fig~\ref{fig:charac}(c)).

For each MZI, we characterised the phase by considering the interferometer as an effective beam splitter with a reflectivity determined by the phase shift and the reflectivities of the four relevant beam splitters.  This was again characterised using the ratiometric technique.  Adjusting the voltage across the thermo-optic element varies the effective reflectivity, as shown in Fig.~\ref{fig:charac}(d).  Fitting this data then provides both an estimate of the zero-voltage phase of the interferometer, and an independent consistency test of the four beam-splitter reflectivities which define each interferometer.  These checks agreed within the measurement uncertainty.

\begin{figure*}
\includegraphics[width=2\columnwidth]{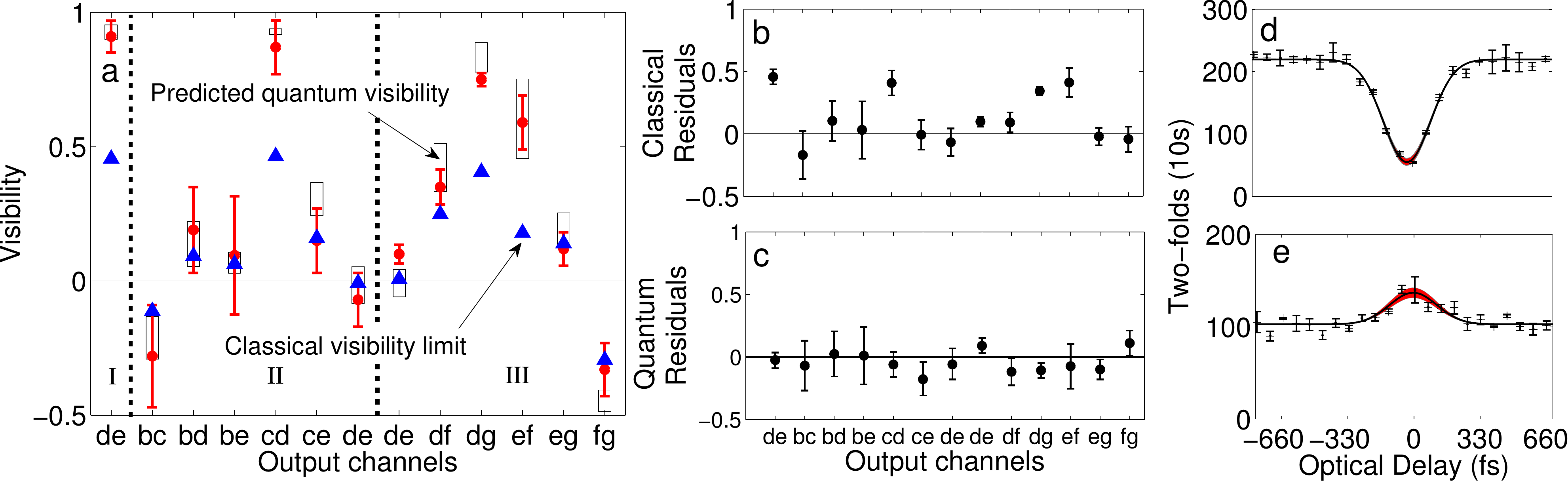}
\caption{
\textbf{Two-photon interference}
\textbf{(a)}
Experimentally measured two-photon interference visibilities (red circles) are compared against
the quantum (clear boxes) and
classical (blue triangles) predictions.  Regions I, II and III contain experiments using input modes \emph{cf}, \emph{cd}, and \emph{df}, respectively. The errors shown on the simulated quantum and classical visibilities
were calculated by Monte-Carlo simulation; errors on the classical visibilities are smaller than the marker size.
\textbf{(b-c)}
The residuals between the measurements and the calculated classical and quantum visibilities.
\textbf{(d-e)}
Example two-fold coincidence counts between output channels \emph{dg} and \emph{fg} when two photons are input into modes \emph{df} and the optical delay of mode \emph{d} is varied.  
}
 \label{fig:2Photon}
\end{figure*}

\subsection*{Demonstrating genuinely quantum operation and circuit complexity}

Having characterised the individual circuit elements classically, we now investigate the operation of the circuit using quantum input light.  We generated three individual input photons from two spectrally factorable down-conversion pair sources, using both photons from one pair and using the other as a heralded single photon (see Methods for details).  Detecting all four photons in this way to give four-fold coincidence measurements reduces the effects of noise terms.

We first study the quantum interference from two-photon inputs to confirm our classical device characterisation and our single-photon indistinguishability.  In these experiments, we injected photons into two of the selected input modes: \emph{cf}, \emph{cd}, or \emph{df} (corresponding to regions I, II and III in Fig.~\ref{fig:2Photon}(a)).  We then measured interference visibilities for all possible coincidence outcomes by varying the timing of one photon using an off-chip optical delay stage (e.g. Fig.~\ref{fig:2Photon}(d)-(e)), and compared the observed values to the quantum (boxes) and classical (triangles) predictions (see Fig.~\ref{fig:2Photon}(a)).  The expected quantum visibilities were calculated directly from the quantum output state predicted by the classically characterised circuit unitary adjusted for the independently measured fidelity between photon pairs. The corresponding classical visibilities were calculated by replacing the input Fock states with phase-averaged coherent states (see Methods).

A chi-squared test verifies that these data are consistent with the quantum predictions whilst in strong disagreement with classical theory~\cite{Ghosh1987}. The likelihood for observing a set of interference visibilities is calculated given measurement uncertainties in the observed interference visibilities and the underlying circuit parameters (see Methods). The residuals from quantum theory give a reduced chi-squared $\chi_r^2=0.9$; a value at least this large will occur with probability $P(\chi_r^2 \ge 0.9) = 0.5$. By contrast, classical theory gives $\chi_r^2=23$ corresponding to $P(\chi_r^2 \ge 23) < 10^{-16}$. Furthermore, the ultimate classical visibility limit of 1/2 is expected and observed to be exceeded by output mode combinations \emph{de}, \emph{cd}, and \emph{dg}. 

The three experiments each used different pair-wise combination of the three chosen input photons, 
allowing the different input photons to be tested individually against each of the other two.  In all three cases, the experiments are in good agreement with the expected quantum results. This suggests that our element-wise characterisation technique is a good predictor of device performance and shows that each of the three photon quantum states has a good fidelity with the other two, a critical factor for observing genuine three-photon interference.

\begin{figure*}[t]

\includegraphics[width=2\columnwidth]{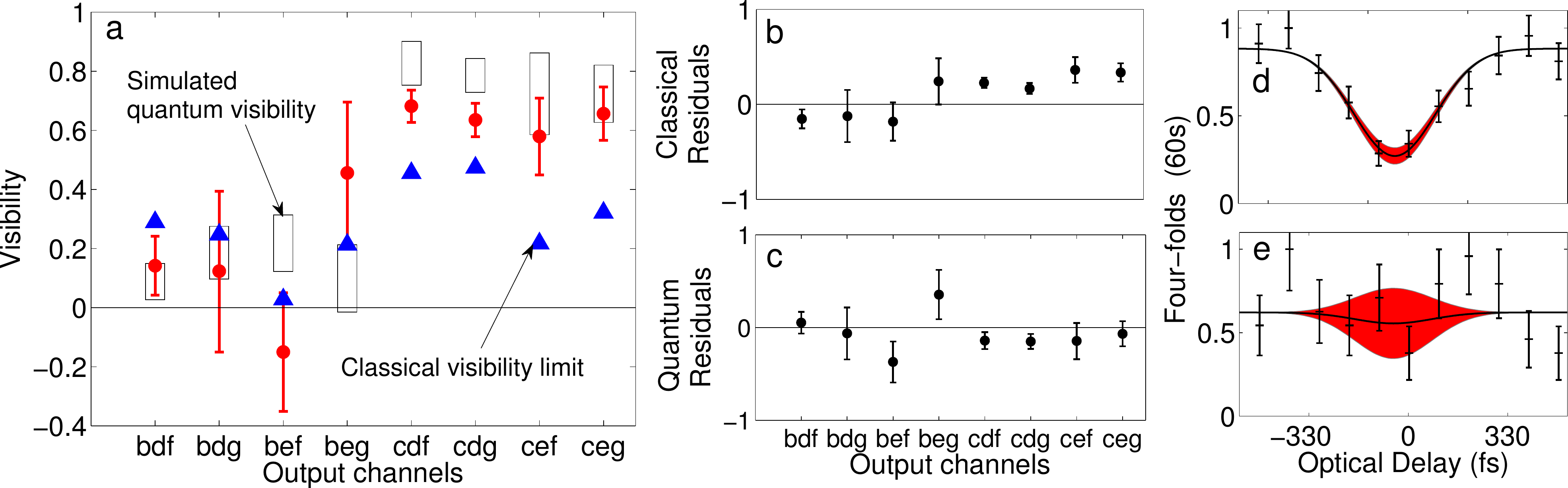}
\caption{
\textbf{Three-photon interference}
\textbf{(a)}
The experimentally recorded three-photon interference visibilities (red circles) are compared against
the predicted quantum (clear boxes) and
classical (blue triangles) results as in Figure~\ref{fig:2Photon}.
\textbf{(b-c)} 
The residuals between our measured and simulated visibilities.
\textbf{(d-e)}
Example four-fold coincidence counts of output channels \emph{cdf} and \emph{bdg} when the input photon in mode \emph{d} is temporally delayed. The shaded area shows the uncertainty in the determined visibility.
 \label{fig:3Photon}
}
\end{figure*}

To demonstrate the complexity of the quantum circuit, we study the higher-order nonclassical interference which arises when the three coupled interferometers are all operating simultaneously and in parallel, each injected with quantum light at the input.  We observe this via the (heralded) three-photon coincidence counts, with three individual input photons coupled into modes \emph{c}, \emph{d} and \emph{f}.  Again detecting the fourth photon enabled discrimination between downconversion events with one photon in each spatial mode ($|111\rangle_\mathrm{cdf}$) from equally likely unwanted noise events with two photons in each of only two input modes ($|022\rangle_\mathrm{cdf}$).  After setting the temporal delay between input modes \emph{c} and \emph{f} to maximize their two-photon interference, we varied the delay for input mode \emph{d} while simultaneously monitoring the eight three-fold coincidence combinations described in Fig.~\ref{fig:3Photon}(a).  We observed average four-photon coincidence rates of around 16~mHz and measured the heralded three-folds continuously for 294~hours, iterating a full scan of the temporal delays each minute.  This method averages out long-term systematic effects, such as drifts in the chip coupling efficiencies and photon source performance~\cite{lanyon2009}, and allows an accurate calculation of the statistical error in the counts. 

As in the two-photon experiments, the observed three-photon quantum interference visibilities agree with quantum predictions based on the individually characterised circuit elements and are completely inconsistent with the equivalent classical predictions. The quantum prediction gives $\chi_r^2=1.5$ with $P(\chi_r^2 \ge 1.5) = 0.2$, whereas the classical prediction gives $\chi_r^2=6$ with $P(\chi_r^2 \ge 6) < 10^{-8}$. Moreover, using a global optimisation routine, we determined the maximum classical interference visibility for any possible circuit parameters with this circuit topology to be 0.59.  This ultimate limit is exceeded by more than one standard deviation by output channel combination \emph{cdf}.  Furthermore, a chi-squared test shows that the circuit parameters which result from this optimisation are strongly inconsistent with our measured values ($P(\chi_r^2 \ge 4.4) = 10^{-8}$).  Thus, only the full quantum explanation can plausibly account for the higher observed visibilities.  We note that Fig.~\ref{fig:3Photon}(a) includes all measured three-fold coincidence combinations, including several that occur very rarely and which therefore lead to large error bars in the measured visibility.  Nevertheless, including all observed data in our analysis, we can exclude classical models with extremely high confidence, despite the low count rates in some channels.

Finally, to verify that the observed interference results from a quantum interaction of all three photons and is not explainable via a bi-separable interaction only, we also simulated the expected quantum visibilities when one of the three photons remains completely distinguishable from the other two at all temporal delays.  These quantum bi-separable explanations are also inconsistent with the data, predicting at most a likelihood of $P<10^{-8}$ for the observed interference signature.

\section*{Discussion}
These results demonstrate genuinely multipartite quantum operation in a next-generation integrated circuit which provides a critical new level of complexity in integrated quantum photonics.  The measurements simultaneously accessed three coupled interferometers, with classical interference at three circuit nodes and nonclassical interference at five circuit nodes.  This experiment also represents the first observation of chip-based, multipartite nonclassical interference which relies on more than two individual photonic inputs.  The three-photon interferences were measured for eight different output combinations and are irreconcilable with classical models.

We develop a practical method to verify successful operation of the device: a loss-independent technique to classically characterise individual circuit parameters and two-photon interference to verify device performance understood by simulations. Using a parameterised characterisation allows identification of poorly fabricated components and freedom to separately simulate individual circuit subsections, such as on-chip state preparation, manipulation and measurement.  Having techniques which successfully predict device performance will be critical as experimental capacities continue to improve, making this demonstration an important step forward.  An alternative scheme has recently been introduced for inferring the overall unitary transformation implemented by a device from a series of classical interference experiments using only the nominal input and output ports~\cite{White2012, Laing2012, meany2012}.  Whilst this does not give access to individual components, it may be useful for cases where only the operation of the device as a whole is of concern. 

In this paper, we highlight the often-unacknowledged fact that minimising losses will be critical for scaling up integrated circuits to the regime where they can no longer be simulated using classical processors.  An array of ongoing work seeks to do so by integrating~\cite{Xiong2011, Martin2010, EcksteinA2011,MartinA2012} and synchronising~\cite{Nunn2012} quantum light sources, as well as developing high-efficiency integrated detectors~\cite{Gerrits2011b, Sprengers2011a}.

Increasing the complexity of integrated photonic devices requires not only an increased number of discrete optical modes but also complex multiphoton quantum interference across all of these modes.  This work has already verified the multiphoton interference necessary for the first nontrivial tests of recently proposed boson sampling problems~\cite{Aaronson2010, Aaronson2011, Rohde2012}.  It also provides the first demonstration of quantum operation of a chip which is sufficiently complex to allow a range of advanced quantum information protocols, such as teleportation and cluster-state generation, to be realised on an integrated platform.

\section*{Methods}
\subsection*{Device fabrication}
\footnotesize{
The waveguide circuit used in this work was fabricated by use of UV-direct write technology on a silica-on-silicon substrate~\cite{smith2009}. The individual waveguides were written by focusing a continuous-wave UV laser (244 nm wavelength) onto the chip which is subsequently moved transversely to the surface normal with computer-interfaced 2D motion control. The UV-writing process enables creation of beam splitters (X couplers) by crossing waveguides at different angles~\cite{kundys2009}. These are much smaller than traditional directional couplers which helps to reduce the effect of propagation loss in more complex circuits. The effective beam splitter reflectivity of these X couplers is primarily governed by the intersection angle of the guides which reduces sensitivity to wavelength, polarisation and temperature fluctuations making them extremely stable over the long experimental durations in this work. The thermo-optic phase shifters utilise a small NiCr electrode (0.35\,$\mu$m $\times$ 50\,$\mu$m $\times$ 2.5\,mm, 0.85\,kOhm electrical resistance) deposited directly over one of the waveguides through which a current can be passed. The temperature-stabilised passive stability of the interferometers with the phase-shifters set to a constant voltage was measured to be less than $1^{\circ}$ over 24 hours.
}

\subsection*{High-brightness multiphoton states on-chip}
\footnotesize{
An 80\,MHz Ti:Sapphire oscillator (Mai-Tai, Spectra Physics) produces 100\,fs pulses at 830\,nm (2.6\,W average power) which are upconverted to 700\,mW of 415\,nm light in a 700$\,\mu\textrm{m}$  $\beta-\textrm{BaB}_2\textrm{O}_4$ (BBO) crystal cut for type-I second-harmonic generation. This blue light is split on a 50:50 beam splitter and used to pump two 8\,mm-long AR-coated Potassium Dihydrogen Phosphate (KDP) crystals phase-matched for degenerate type-II collinear parametric down-conversion. We optimise the collection optics and spatial mode-matching to achieve a coincidence count rate of 160\,kHz on each crystal with a raw heralding efficiency of 28-30\,\% without any filters. The source is designed to be spectrally factorable~\cite{MosleyPJ2008hgu} which improves the heralding efficiency we can achieve when interference filters (Semrock, $\Delta\lambda=3$\,nm) are used to match the bandwidths of the broad and narrowband daughter photons. With the filters in place we achieve a four photon coincidence rate of 20\,Hz and two-photon fidelities of 0.99 (narrowband-narrowband) and 0.96 (narrowband-broadband).

Three of the photons were coupled into polarisation maintaining (PM) fibers and launched into the waveguide circuit using a butt-coupled PM v-groove array. Index matching oil is applied between the fiber array and the chip to reduce reflection losses and a 6-axis piezo-controlled alignment stage provides all the degrees of freedom necessary to achieve optimal simultaneous coupling into all six input modes. The piezo-driven axes were operated in closed-loop mode to maintain this coupling throughout the experiment. An identical set-up is used on the output to a achieve maximal coupling from the chip into the single photon counting modules. A home-built FPGA-based logic unit records all desired coincidence counts simultaneously.

}
\subsection*{Predicted visibilities}
\footnotesize{
The simulated visibilities in our interference experiments were calculated by first simulating the complete quantum output state, $|\psi_\mathrm{out}\rangle$, using the characterised circuit unitary, $\mathbf{U}_\mathrm{circ}$. The intensity cross-correlation functions at zero and infinite temporal delay were then used to find an interference visibility. 

When photons are launched into modes $\{a_i\}$ of a linear optical circuit the intensity cross-correlation between output modes $\{b_i\}$ at zero temporal delay is:
\begin{equation}
\Gamma^{(0)}_{{\{b_i\}}} = \left\langle{\psi_\mathrm{out}^{\{a_i\}}}\right|\left(\prod_{b_i}{\mathrm{\hat{I}}_{b_i}}\right)\left|{\psi_\mathrm{out}^{\{a_i\}}}\right\rangle,
\end{equation}
where the intensity operator on mode $i$ is $\hat{\mathrm{I}}_i=\creop{b}_i\hat{\mathrm{b}}_i$. Suppose the photon in mode $a_d$ undergoes a temporal delay then the total output state is now a classical mixture:
\begin{equation}
\rho_\mathrm{out} =
\proj{\psi_\mathrm{out}^{\{a_i,i\neq d\}}}{}{\psi_\mathrm{out}^{\{a_i,i\neq d\}}}
+\proj{\psi_\mathrm{out}^{(a_d)}}{}{\psi_\mathrm{out}^{(a_d)}}.
\end{equation}
The intensity cross-correlation function at infinite delay is then,
\begin{equation}
\Gamma^{(\infty)}_{\{b_i\}} = \mathrm{Tr}\{\rho_\mathrm{out}\prod_{b_i}{\mathrm{\hat{I}}_{b_i}}\},
\end{equation}
and the visibility of the interference pattern between the $n^{th}$-order output coincidence counts is then given by
\begin{equation}
V_\mathrm{quant} = \frac{\Gamma^{(\infty)}_{\{b_i\}} -
  \Gamma^{(0)}_{\{b_i\}}}{\Gamma^{(\infty)}_{\{b_i\}}}.
\end{equation}
The corresponding classical visibility of this interference pattern is given by injecting three equal amplitude coherent states of mutually randomised phase into modes $\{a_i\}$. This ensures we mimic independent sources of light which will have no first-order correlation as required to compare against Hong-Ou-Mandel-type quantum interference. Coherent states represent the classical state which have the highest interference visibility, ensuring the bound we calculate is an upper limit. The resulting output vector of complex amplitudes is:
\begin{align}
\mathbf{e}_\mathrm{out} &=\mathbf{U}_\mathrm{circ}\mathbf{e}_\mathrm{in}\\
&=\mathbf{U}_\mathrm{circ}\left(\begin{matrix}e^i\theta_1\\ \vdots \\e^{i\theta_n}
  \end{matrix}\right),
\end{align}
where $\mathbf{e}_\mathrm{out}$ is the vector of time-independent
electric fields in each of the output modes and similarly for
$\mathbf{e}_\mathrm{in}$.
The phase-averaged intensity cross-correlation function is then
\begin{equation}
\Gamma'^{(0)}_{\{b_i\}} = \frac{1}{(2\pi)^n}\int_0^{2\pi}\!\!\hdots\!\!\int_0^{2\pi}\prod_{b_i}{|(\mathbf{e}_\mathrm{out})_{b_i}|^2}\,d\theta_1\!\!\hdots d\theta_n.
\end{equation}
The classical cross-correlation function at infinite delay is calculated in a similar
manner, taking incoherent sums between the delayed and non-delayed photons,
\begin{equation}
\Gamma'^{(\infty)}_{\{b_i\}} = \frac{1}{(2\pi)^n}\int_0^{2\pi}\!\!\hdots\!\!\int_0^{2\pi}\prod_i{
\left(|(\mathbf{e}_\mathrm{out}^{\{a_i,i\neq d\}})_\mathrm{b_i}|^2 +
  |(\mathbf{e}_\mathrm{out}^{(a_d)})_\mathrm{b_i}|^2\right)}\,d\theta_1\!\!\hdots d\theta_n.
\end{equation}
From which we calculate the classical interference visibility,
\begin{equation}
V_\mathrm{class} = \frac{\Gamma'^{(\infty)}_{\{b_i\}} -
  \Gamma'^{(0)}_{\{b_i\}}}{\Gamma'^{(\infty)}_{\{b_i\}}}.
\end{equation}
}

\subsection*{Chi-squared test} 
To estimate the likelihood of observing a particular set of \emph{m} interference visibilities $\mathbf{v}$, we construct the probability density function
\begin{equation} f(\mathbf{v})\sim \int\left(\prod_i^n{\exp{\left(\frac{-\alpha_i^2}{2s_i^2}\right)}}{\prod_i^m{\exp{\left(\frac{-(v_i-w_i(\mathbf{\alpha}))^2}{2\sigma_i^2}\right)}}}\right){d^n}\mathbf{\alpha},
\label{eq:pdf}
\end{equation}
where $w_i(\mathbf{\alpha})$ is the calculated $i^{\mathrm{th}}$ visibility based on the set on $n$ circuit parameters $\mathbf{\alpha}$, $\sigma_i$ is the measurement standard deviation in the observed visibility $v_i$, and $s_i$ is the measurement standard deviation of the characterised circuit parameter $\alpha_i$. We approximate each $w_i(\mathbf{\alpha})$ to be linear in $\alpha$, \begin{equation}
w_i(\mathbf{\alpha}) \approx w_i^0 + \sum_j^n{\frac{\partial w_i}{\partial \alpha_j}\alpha_j}, \end{equation}
which we verify to be accurate to within 0.02 over the range $\pm 2 s_j$. The resulting multidimensional Gaussian integral yields an analytic solution with an exponent quadratic in visibility residuals. We analyse our data in a basis $x_j = \sum_i c_{ij}(v_i-w^0_i)$ so that $f(x)$ is factorable:
\begin{equation} f(\mathbf{x})\sim \prod_i^m\exp\left(-\lambda_ix_i^{2}\right).
\end{equation}
The visibilities $\mathbf{x}$ are thus statistically independent and a typical chi-squared test, $\chi_r^2 = 1/m \sum_i^m{\lambda_ix_i^2}$, can be applied. If we instead errantly assume that the uncertainties in our simulated visibilities are uncorrelated, we calculate a chi-squared approximately 0.3 lower than that reported for the quantum analysis of our experiments. 

\section*{Acknowlegements}
This work was supported by the EPSRC(EP/C51933/01), the EC project Q-ESSENCE (248095), the Royal Society, and the AFOSR EOARD. XMJ acknowledges support from NSFC (11004183) and CPSF (201003327). NKL is supported by an EC Marie Curie fellowship. MB is supported by a FASTQUAST ITN Marie Curie fellowship.


\end{document}